# Adoption and Use of LLMs at an Academic Medical Center

Nigam H. Shah, Nerissa Ambers, Abby Pandya, Timothy Keyes, Juan M. Banda, Srikar Nallan, Carlene Lugtu, Artem A. Trotsyuk, Suhana Bedi, Alyssa Unell, Miguel Fuentes, Francois Grolleau, Sneha S. Jain, Jonathan Chen, Devdutta Dash, Danton Char, Aditya Sharma, Duncan McElfresh, Patrick Scully, Vishantan Kumar, Clancy Dennis, Connor O'Brien, Satchi Mouniswamy, Elvis Jones, Krishna Jasti, Gunavathi Mannika Lakshmanan, Sree Ram Akula, Varun Kumar Singh, Ramesh Rajmanickam, Sudhir Sinha, Vicky Zhou, Xu Wang, Bilal Mawji, Joshua Ge, Wencheng Li, Travis Lyons, Jarrod Helzer, Vikas Kakkar, Ramesh Powar, Darren Batara, Cheryl Cordova, William Frederick III, Olivia Tang, Phoebe Morgan, April S. Liang, Stephen P. Ma, Shivam Vedak, Dong-han Yao, Akshay Swaminathan, Mehr Kashyap, Brian Ng, Jamie Hellman, Nikesh Kotecha, Christopher Sharp, Gretchen Brown, Christian Lindmark, Anurang Revri, Michael A. Pfeffer

Corresponding author: Nigam H. Shah, nigam@stanford.edu
Stanford Medicine, Technology and Digital Solutions

# Abstract

While large language models (LLMs) can support clinical documentation needs, standalone tools struggle with "workflow friction" from manual data entry. We developed ChatEHR, a system that enables the use of LLMs with the entire patient timeline spanning several years. ChatEHR enables automations - which are static combinations of prompts and data that perform a fixed task - and interactive use in the electronic health record (EHR) via a user interface (UI). The resulting ability to sift through patient medical records for diverse use-cases such as pre-visit chart review, screening for transfer eligibility, monitoring for surgical site infections, and chart abstraction, redefines LLM use as an institutional capability. This system, accessible after user-training, enables continuous monitoring and evaluation of LLM use.

In 1.5 years, we built 7 automations and 1075 users have trained to become routine users of the UI, engaging in 23,000 sessions in the first 3 months of launch. For automations, being model-agnostic and accessing multiple types of data was essential for matching specific clinical or administrative tasks with the most appropriate LLM. Benchmark-based evaluations proved insufficient for monitoring and evaluation of the UI, requiring new methods to monitor performance. Generation of summaries was the most frequent task in the UI, with an estimated 0.73 hallucinations and 1.60 inaccuracies per generation. The resulting mix of cost savings, time savings, and revenue growth required a value assessment framework to prioritize work as well as quantify the impact of using LLMs. Initial estimates are $6M savings in the first year of use, without quantifying the benefit of the better care offered. Such a "build-from-within" strategy provides an opportunity for health systems to maintain agency via a vendor-agnostic, internally governed LLM platform.

# Introduction

The rapid rise of large language models (LLMs) in healthcare has created both opportunities and challenges for health systems. While generative AI has shown promise in clinical communication, documentation support, and knowledge retrieval, most adoption to date is in the form of commercial tools. This creates challenges related to privacy, workflow misalignment, safety, and loss of institutional agency in shaping how LLMs are used in patient care [1]. Other work has also shown that LLM pilots frequently fail when organizations do not invest in change management, real-world evaluation, and user enablement, underscoring the need for intentional deployment strategies in healthcare settings [2].

On January 29th, 2024, we launched SecureGPT, a web application enabling the use of patient records and other protected information with multiple LLMs in our organization's secure cloud infrastructure. As of this writing, 18 different LLMs from OpenAI, Anthropic, Google, Meta and DeepSeek are available in SecureGPT [3]. Initial usage patterns correlated with known LLM usage trends [4,5] but also revealed a key limitation: as a standalone web application, SecureGPT required users to copy and paste chart content, creating workflow friction and limiting adoption for longitudinal summarization and other context-heavy tasks. Meanwhile, frontline clinicians face rising documentation burden and spend substantial time synthesizing longitudinal patient records [6]—an area where generative AI could provide meaningful relief [7].

Much of the evaluation of LLMs in medicine remains narrowly focused on test accuracy or performance on synthetic question-answering tasks [4], which provides limited insight into real-world workflow impact, user adoption, or patient-care benefit. Our efforts to assess LLM performance in the real-world in a task-specific manner found that most models performed well in clinical document generation and patient communication tasks while struggling with clinical decision support[8]. In addition, a multi-institution survey on physician perspectives on LLM use found that institutionally approved access was viewed as necessary and most respondents (94%) felt training was necessary to ensure safe clinical use [9].

From these findings, we concluded that safe and effective use of LLMs in clinical care requires: real-time access to the longitudinal medical record inside the LLM context window, (2) direct embedding into clinical workflows, and (3) continuous evaluation, coupled with user training and feedback. Therefore, we built ChatEHR, a three-part system in which the longitudinal medical record of a patient is inserted in real-time into the context window of a LLM (*ChatEHR-platform*) and can be accessed both via a programmatic interface (*ChatEHR-API*) as well as an interactive user-interface (*ChatEHR-UI*) to support a range of tasks. This endeavor was possible because of an established responsible AI lifecycle—bringing together the executive leaders across Stanford Medicine—to guide and oversee AI solutions across the enterprise.

We describe our end-to-end approach to deploying GenAI in an academic medical center (AMC) and community hospital, including system design, evaluation, rollout, monitoring, and early outcomes, with the goal of offering a replicable and responsible setup for clinical use of LLMs.

# Methods

## ChatEHR Platform

The ChatEHR Platform (described further by McElfresh et al [10]) provides a set of capabilities that safely connect LLMs with real-time clinical data at the point of care. The Platform (Figure 1) has three capabilities: 1) Data Orchestration, which allows extracting data elements by time range, type of encounters, visit location and other attributes that define the context of the patient record to include when responding to a given prompt [11]; 2) Context Management, which allows queries, combined with the patient data, to be processed in a token efficient manner; and 3) LLM Routing, which chooses a given LLM based on the token count, and, if the count exceeds the model's context window, splits the context into chunks for parallel processing. The majority of the patient data is processed in Fast Health Interoperability Resources (FHIR) format, but some content—such as referral records, information about organ transplants—come in a non-FHIR format and are processed in their native form.

The Platform supports three types of usage via application programming interfaces (APIs): 1) For developers, who access the Platform for rapid exploration; 2) For scheduled or triggered processes that automate rote tasks for a group of patient records; and 3) A user interface (UI) application, which is a chat interface embedded within the electronic health record (EHR).

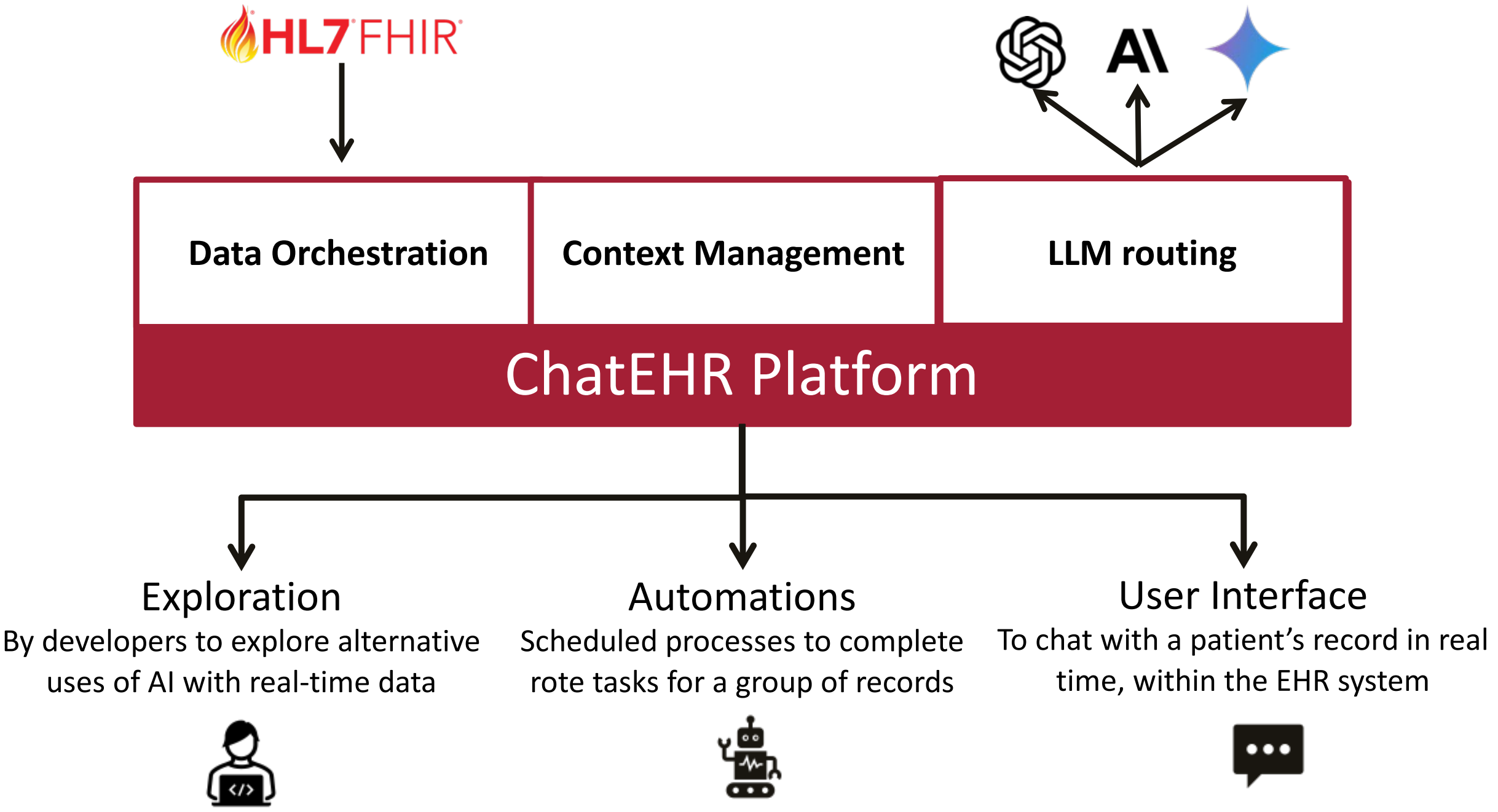

Figure 1: The ChatEHR Platform's three capabilities (Data Orchestration, Context Management, and LLM routing) that support three types of usage via APIs. Developers explore alternative uses of AI

capabilities with real-time data, automations complete rote tasks for a group of records, and the UI allows a conversation chat with a single patient's record within the EHR.

## Evaluation Framework

Current state-of-the-art methods to evaluate GenAI systems is via benchmark-based evaluation. For example, MedHELM introduced a task-based extensible evaluation framework for assessing LLM performance for medical use. It provides a clinician-validated taxonomy of 121 medical tasks organized into 5 categories and 22 subcategories along with 37 benchmarks covering clinical decision support, note generation, patient communication, research assistance, and administrative workflows that mirror real-world scenarios. The initial MedHELM results informed the choice of LLMs to include in the ChatEHR Platform. However, there is no single benchmark, or set of benchmarks, that can anticipate all subsequent uses of an LLM-powered system pre-deployment.

When the task to be done is pre-defined, as is the case for an automation, we create a representative dataset for testing. These datasets also serve as a continuously updated reference standard for post-deployment monitoring *system integrity* (whether automation pipelines run without errors), *performance* (whether LLM outputs remain accurate when compared with the gold standard), and *impact* (whether automations deliver the intended clinical or operational benefit).

When the task a user will do is not known in advance, e.g. in the UI, we combine quantitative metrics (such as hallucination rates) with ongoing qualitative assessments (such as relevance). We rely on processing the interaction logs into categories of tasks being done along with requesting user feedback on the quality of the response as well as the specifics of the error made. These interaction logs are then analyzed to identify common tasks (such as summarization) and corresponding analyses (such as estimating hallucination rates) to quantify the rate of errors.

## Automations

Automations are static combinations of prompts and code scripts that perform a fixed action from a wide variety of data sources. Automations can be triggered, batch-executed, or called on-demand. For example, the repeated submission of instructions to a LLM for processing a longitudinal medical record—such as creating a pre-visit patient summary or determining transfer eligibility—Is an automation.

### Pilot Study

To identify a repeatable process to create automations, we piloted two use cases between February and October 2024. These pilots also established the feasibility and limitations of connecting EHR data to LLMs to address targeted problems.

The first pilot summarized the pre-visit history of present illness for providers in one Otolaryngology clinic over a six-week period. The automation summarized symptoms, medications, imaging results, and referral reasons using one year of notes (such as progress, orders, history & physicals (H&Ps), procedure, consult) for 29 patients seen by 3 physicians. The second pilot delivered a shift summary draft to nurses by extracting key data charted throughout the shift, including flowsheets, medications, and measurements. The prompt was developed by defining contents for an ideal note for patients with acute chest pain based on flowsheet, medication, and result data from the patient record within the shift timeframe and evaluated by 6 nurses for correctness, completeness, conciseness, and harm.

During the pilot, we also constructed gold standard labeled datasets consisting of lists of patient's unique identifiers, date ranges for the portion of the record to be used, questions/prompts used, and an expert-curated response. In addition to the initial evaluation, having these benchmark datasets allows us to monitor the quality of the LLM responses over time.

### Monitoring and Evaluation

For each new automation, we curate a task-specific, gold standard dataset that includes a patient identifier, the corresponding record used during feasibility assessment, the prompt, and clinician-adjudicated labels. Doing so allows initial benchmarking of performance. In addition, we rerun each automation's associated benchmark whenever specific LLMs are updated or deprecated and recompute evaluation metrics to guide decisions about how to update or swap each automation's underlying LLM.

For *system integrity* monitoring, all automation jobs log request-level telemetry such as latency, token-usage, and error codes. For each automation, we track the number of patient records processed and the number and types of inference errors (such as API errors or timeouts). We also periodically re-extract a subset of patients in the gold standard dataset to compare the new data to the original benchmark—allowing us to detect failures in data retrieval or changes to upstream documentation.

For *performance* monitoring, we collect structured, in-workflow feedback from clinicians on the automation output—whether they agree or disagree with the LLM's recommendations or identified errors in its reasoning—which allows us to track quality over time and "append" these adjudications into the benchmark dataset for each automation.

For *impact* monitoring, we collect metrics about what users do with the outputs and how often they use them. Depending on the task, we track: 1) action rates e.g. the proportion of flagged patients who received the recommended follow-up, such as referral to a specialty service; 2) workload and documentation metrics, such as reductions in manual chart review time, use of generated text in notes, or changes in chart abstraction time; and 3) basic usage and engagement metrics such as how often summaries or reports are opened, viewed, and copied. Additionally, because automations rely on LLM APIs with per-request pricing, we also track runtime cost as an additional impact metric for assessing benefit.

#### Value Assessment

To ensure consistency in value assessment for an automation, our finance team developed a structured framework, implemented as a custom-built tool, that uses standardized estimates for cost savings, time savings, and revenue growth to produce financial projections. First year projections reflect conservative assumptions including partial rollout, limited utilization, and learning curve effects; while steady-state projections approximate year three adoption with stable workflows and optimized use. Benefits are organized into three categories: 1) Cost Savings: direct monetary reductions in operational expenses; (2) Time Savings: decreased time spent on manual, repetitive, or administrative tasks, returning capacity to the workforce (these savings are considered “soft” as the time saved is repurposed rather than eliminated); and 3) Revenue Growth: Incremental revenue driven by new workflows, increased volume, improved throughput, or enhanced margins.

### User Interface

The UI offers a ‘chat’ interface like consumer web applications such as ChatGPT. The UI is primarily used for open-ended conversations but can also be used to explore opportunities for automations by trying out different queries in a no-code setting.

#### Pilot Study

Pilot testing of the UI began in January 2025 with a 15-user alpha group and expanded to over 150 participants by the end of the pilot in June 2025. Users tested the UI and participated in focused feedback rounds, providing detailed input on features, usability, and readiness for broader rollout.

Feedback was collected through two structured surveys and supplemented with small-group and individual sessions for deeper discussion. We also conducted 15 semi-structured interviews with clinicians and administrative staff, including physicians, nurses, case managers, and revenue cycle leaders, focused on workflow pain points and desired functionality. In addition, a nursing shared leadership group (n=40) was introduced to the tool through screenshots and demo videos and surveyed for their input.

Feature requests and pain points were triaged by the core team based on perceived clinical impact and feasibility of implementation. Enhancements, such as integration of additional data sources (e.g. real-time data feeds, external health information exchange data), were made in response to feedback during the pilot testing. A dedicated Microsoft Teams channel provided day-to-day outreach, troubleshooting, and education. Significant themes and recurring questions identified in focus groups or chats were addressed, both through new UI feature development (e.g. allowing selection of time range and data sources) and posting on the channel (e.g. system performance issues). The system prompt was finalized as described in Appendix A.

Based on the top tasks that were reported on[4], observed in SecureGPT usage[5], and mentioned in user surveys, we concluded that evaluating the quality of summaries produced using the UI would be a necessary pre-deployment performance test. While evaluation of a 'summary' remains an active research area, the most common approach to evaluation is via calculating a FactScore [12], which breaks LLM-generated prose into a series of atomic facts and computes the percentage of atomic facts supported by a reliable knowledge source—in our case the full underlying patient record. To evaluate the quality of summaries we used a strategy adapted from VeriFact[13] to identify 'unsupported claims' (described in Appendix B). An unsupported claim can be either a *hallucination* which is a statement in the summary that cannot be found in or supported by the retrieved source text (e.g., mentioning a procedure that is not documented in the record) or an *inaccuracy* which is a statement in the summary that explicitly contradicts information present in the source text (e.g., reporting a lab value of 5.0 when the record states 3.5). We analyzed a one-month sample of clinician-ChatEHR conversations from July 1 to July 31, 2025 collected during the pilot testing of the UI.

### Monitoring and Evaluation

The UI, shown in Figure 2, was made broadly available on September 9, 2025 to all physicians and advanced practice provides (APPs) totaling about 7000 users at Stanford Medicine. Because the tasks for which the UI can be used is unbounded, users were asked to view a training video that reviewed the key features of the UI, shared tips for prompting, and stressed the need to verify LLM outputs before using them for decision making.

Like automations, we monitor the UI for system integrity (in terms of response times, errors and time outs, number of daily active users, daily and weekly sessions and data types used), performance (in terms of the most common tasks, the user-reported errors, and analyses for quality of summaries), and impact (in terms of clinical or operational benefit as well as LLM API cost).

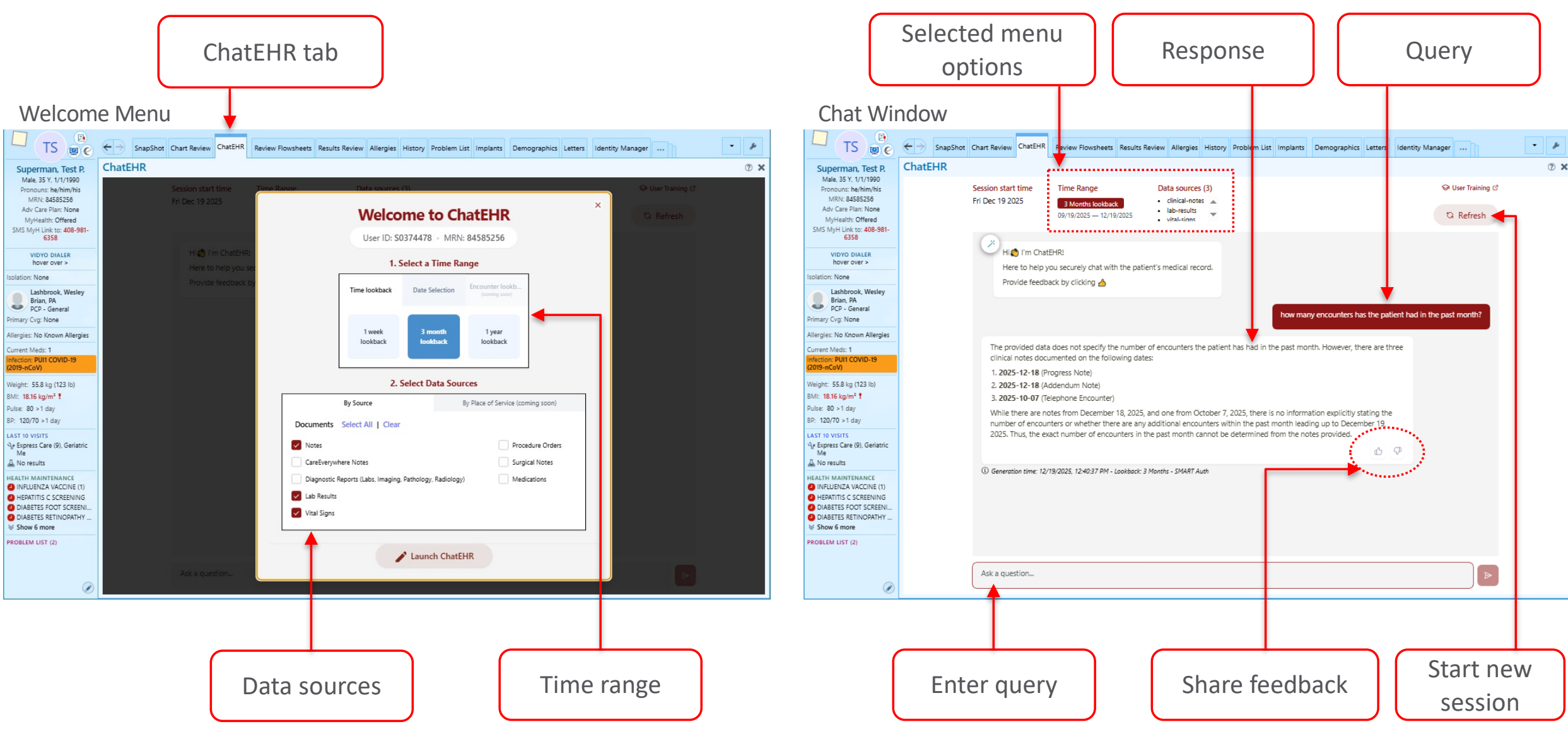

Figure 2: When users open a ChatEHR tab in the EHR, they first see a welcome menu (left) to select which patient data to load into context for the LLM, by selecting data sources and a time range. After launching ChatEHR, they see a chat window (right) where they can submit queries and receive responses, provide feedback (thumbs up/down), or start a new session by returning to the welcome menu.

The data necessary for monitoring system integrity and impact is collected via logging metadata on each user session and analyzed in dashboards in DataBricks. The data necessary to monitor performance is obtained via processing the session logs from September 9 to December 1, 2025. The session logs are processed to identify common tasks (such as summarization), user feedback on the quality of the responses, and analyses to quantify the rate of errors (such as estimating unsupported claims rates) as well as to identify the factors associated with those errors.

To identify common tasks, we categorize the tasks performed in the UI by combining linguistic classification with clustering-based medical intent categorization as described in Appendix C. To validate this categorization, clinicians assessed 100 user queries for agreement with the automated categorization. User feedback (thumbs up/down) is analyzed to obtain task specific error rates.  To monitor the quality of summaries, we estimate the rate of unsupported claims in summaries applying the same method as before (Appendix B) on a 10% sample of sessions from September 9 to December 1, 2025.

### Value Assessment

Applying the lens of cost savings, time savings, and revenue growth to an interactive chat interface used by a diverse group of users is difficult. Therefore, we approximate the benefit using number of users, minutes saved per user, along with a median hourly salary for the user's time to arrive at a potential time savings estimate. We do not attempt to attribute revenue growth to the use of the chat interface. The quantification of the 'value' of the benefit incurred to patients is also left unaccounted.

### Governance and Team Design

The adoption of AI at Stanford Health Care (SHC) is a multi-pronged effort to enable traditional predictive modeling initiatives, such as clinical risk prediction, while also capitalizing on the transformative potential of generative AI. To ensure consistency across efforts, we established an enterprise Responsible AI Lifecycle (RAIL) in 2023 that set guidelines for risk tiering, the required assessments to ensure Fair Useful Reliable Models (FURM) [14], and subsequent monitoring of system integrity, performance, and impact[15]. RAIL codifies institutional workflows for how AI solutions are proposed, evaluated, validated, and monitored. A public facing website outlines how we follow RAIL to serve our patients, providing faster, safer, and more personalized care for all[16]. This includes a structured ethics assessment, when applicable, incorporating stakeholder input on perceived benefits, risks, and burdens of the tool, equity considerations, and preferences for governance and patient notification[17]. In addition, we aligned RAIL with our product development lifecycle and DevOps release cycles to translate prototypes into validated, scalable solutions, establishing a disciplined approach to governance, testing, release, and monitoring of LLM use.

This integrative approach was made possible by embedding the data science team, which includes software engineers, product managers, and data scientists with direct access to academic researchers, in the IT organization[18]. As a result, the team members have direct access to personnel who maintain network security, make integrations across multiple IT systems including the EHR, design enterprise software architecture and manage cloud resources. In addition, the team includes members with operational and clinical experience as well as faculty and students who can make technical recommendations, such as design evaluation and monitoring plans, based on state-of-the-art frameworks such as FURM[14] and MedHELM[8] as well the latest research such as VeriFact[19]. Because of this set up, what began as a sandbox in April 2023 was conceptualized as ChatEHR in November 2023, underwent multiple prototypes by mid-2024, leading to a pilot in early 2025 that scaled to 150 clinical users by August 2025, and a health system-wide roll out by September 2025.

## Results

The ability to sift through patient medical records to interpret and summarize the information is an all-too-common manual chore in current medical practice. ChatEHR enables the use of LLMs for such tasks by retrieving the entire patient timeline spanning several years and making it available for use in the context window of a LLM to drive automations and to enable interactive use via the UI.

## Automations

Automations, such as creating a pre-visit patient summary or determining transfer eligibility, are effective when the instructions (also called prompts) to the LLM do not change frequently, the data elements on which the instruction is to be applied are fixed, and performance is verifiable via a truth-set. The most valuable automations are integrated into specific action-oriented workflows to drive review, decisions, and next events.

### Pilot Study

In the pre-visit history summarization in the Otolaryngology clinic, two-thirds of generated summaries were determined usable with 95.0% agreement with physician consensus as reported by Liang et al[20]. In the pilot delivering a shift summary draft to nurses, one-eighth of the generated summaries were determined useable, requiring significant context and prompt refinements. Both projects confirmed the need for custom data retrieval, including both real-time access as well as to support data types, such as notes, labs and medications. The pilot efforts underscored the importance of having subject matter experts refine prompts in a close feedback loop within the clinical or operational workflow. Lastly, delivering the output into existing workflows was necessary to optimize usage and impact.

Overall, the two pilots established a repeatable process to build automations, comprised of identifying the workflow trigger, gathering context data, standardizing the set of prompts, creating a truth-set for evaluation in partnership with domain experts, delivering the output integrated into the user workflow, capturing human-in-the-loop feedback, and monitoring.

After the pilots, we implemented seven automations that address tasks such as timely eligibility determination for patient transfers/programs, detection of suspected events, and extraction of essential data.

### Monitoring and evaluation

The seven automations in routine use are monitored in terms of system integrity, performance, and impact as described in the methods. Examples of metrics used to monitor each automation are provided in Table 1.

*System integrity:* System integrity monitoring is performed by observing total executions, inference errors, token consumption, and end-to-end latency for each job. Even across a high volume of inference runs (over 7,500 jobs), jobs typically complete successfully with low absolute error counts. Latency is variable across automations but typically increases with record size and task complexity (e.g. SSI surveillance operates at the largest token volumes, and SCM screening uses a reasoning model).

*Performance:* Performance is captured via in-workflow human-in-the-loop feedback when possible. For example, in-workflow feedback from the hospice automation (81% positive) and surgical co-management automations (77.9% positive) are deemed as positive performance when a clinician agrees that a patient may be (or may become) suitable for a consultation. For the SSI surveillance automation—which is designed as a high sensitivity, low specificity screen—performance is deemed positive when a trained reviewer agrees that a patient experienced an SSI, with a target rate of 10%. Some automations, such as transfer to a lower acuity site, have NA performance metrics because they were implemented before in-workflow feedback was consistently captured.

*Impact:* Like performance metrics, impact metrics are automation-specific and tied to an automation's intended downstream actions or human workload reduction. Action-rate tracking shows that 20% of user-confirmed patients flagged by ChatEHR are referred to inpatient hospice care and 50% of confirmed surgical co-management candidates are seen by the service. For automations wherein downstream actions are difficult to observe, impact is primarily assessed through qualitative user feedback (marked as NA in Table 1).

| | System Integrity | | | | Performance | | | Impact | | |
|---|---|---|---|---|---|---|---|---|---|---|
| **Automation** | **Total Executions** | **Errors** | **Tokens Sent** | **Latency** | **Positive Feedback** | **Negative Feeback** | **Structured Feedback** | **Action Rates** | **Workload and Documentation** | **Usage and Engagement** |
| Identify patients eligible for transfer to a lower acuity site | NA | | | | | | | | | |
| Review clinical criteria for inpatient hospice | 215 jobs; 36,643 patients | 5 | 20,509 per patient 3.5M per job | 872s | 81.25% | 18.75% | NA | 20% of LLM-flagged, user-confirmed patients referred to hospice | NA | NA |
| Identify patients for surgical co-management | 1,875 jobs; 4,762 cases | 73 | 120,947 per case | 1,451s | 77.90% | 22.10% | NA | 50% of LLM-flagged, user-confirmed cases seen by SCM | ~60 min/day physician time savings | NA |
| Surveil patients for surgical site infections | 115 jobs; 115 cases | 2 | 7,857,661 per case | 608.5s | 26% | 74% | NA | NA | ~43.5% reduction in monthly case review workload | NA |
| Streamline donor offer prep for the transplant team | NA - In early pilot phase | | | | | | | | | |
| Extract key data for orthopedics referrals | 2,709 jobs | 26 | ~426 per job + binary document | ~517s | NA | NA | NA | NA | ~100 Min a day savings | 100% Opened Messages |
| Assist with reviewing hospital account letter of agreement | 2,910 jobs | 23 | ~1,560 per job + binary document | ~13s | 93.98% | 6.02% | NA | NA | NA | NA |

Table 1: The table lists the task accomplished by a given automation and the corresponding system integrity, performance and impact metrics. NA denotes not available.

## Value Assessment

Quantifying the benefit, particularly of time savings, remains challenging because saved minutes often occur in fragments that cannot be used to perform other tasks—as seen in the use of ambient scribe tools[21]. For example, the Sequioa transfer automation avoids manual review of approximately 120 charts per day, equating to roughly four hours of case management time, which do not translate directly into additional work completed.

As described in the methods, standardized estimates for cost savings, time savings, and revenue growth are used to produce financial projections (Table 2). To operationalize the value framework, we estimate benefit at three stages in an automation's life cycle—intake, implementation, and monitoring—as illustrated below:

- Intake Stage (Transfer to our community hospital Automation)
  This automation, which is in consideration for implementation and accelerates transfers to our community hospital, is estimated to have $2.3M-$3.2M annual impact from gaining 5 inpatient beds per day (equivalent to 5,220-7,330 bed days annually) and associated new admissions. Labor savings contribute minimally (~$75K).
- Implementation Stage (Inpatient hospice eligibility Automation)
  Designed to identify patients appropriate for hospice care, this automation demonstrated significant time savings (~6,570 hours/year, equivalent to 3 FTEs). However, impact was difficult to quantify due to the absence of a dedicated hospice unit and limited ability to monetize by freeing up beds.
- Monitoring Stage (Transfer to Sequoia patient care unit Automation)
  This in-use automation, which streamlines eligibility screening for transfers to a lower-acuity site, frees up inpatient capacity for higher-acuity cases. While the time saved and associated labor savings are modest (~$100K), there is $2.4-$3-3M annual revenue growth from 1,700 transfers per year.

At the intake stage, projections rely on directional estimates and conservative adoption scenarios, while monitoring stage estimates are based on observed utilization and operational metrics. Strategic benefits, such as improved patient flow, capacity planning, and quality outcomes remain challenging to quantify.

| **Automation** | **Problem** | **Solution** | **Value Assessment** |
|---|---|---|---|
| Identify patients eligible for transfer to a lower acuity site | The Sequoia patient care unit was opened to help resolve capacity constraints and limited bed availability at the Palo Alto main campus. | Streamline eligibility assessments for transfers from Palo Alto ED and Inpatient, reducing manual effort and improving workflow efficiency. | Each day the automation narrows 150 patient charts to roughly 30 highly probable cases reducing screening time. At 2 min/chart this is a potential time savings of 4 hours/day. Appropriately transferring care to |

| | | | lower acuity setting is anticipated to be worth $2-3 Million/year. |
|---|---|---|---|
| Review clinical criteria for inpatient hospice | Identifying appropriate, timely candidates for hospice care is difficult. | Review clinical notes and add flagged patients to a list for Hospice staff to vet and reach out to the corresponding care teams. | The automation narrows ~600 patient charts to a potentially eligible list of 10-40. This automation unlocks a review that would not be possible with current resourcing, at 2 min/chart, this is equivalent to 18 hours/day. The impact on increased appropriate and timely referrals is being monitored for the recent enterprise rollout. |
| Identify patients for surgical co-management | Manual identification of suitable surgical patients for co-management requires significant resources and diverts clinicians from direct patient care, care coordination, and research | Identify patients suitable for surgical co-management and provide the patient list into existing workflows for hospitalists. | Across three surgical areas, the automation saves approximately 1 hour/day, allowing hospitalists to spend more time at the bedside. |
| Surveil patients for surgical site infections | Identifying infections following surgeries requires extensive manual chart review for quality monitoring. This process, while time-consuming, is critical for reporting requirements. | Streamline review by flagging patients with suspected associated infections, allowing the reviewers to focus on the most relevant charts. | The automation reduces manual review effort, streamlines surveillance, and supports adherence to reporting standards. The system reduces the number of patients charts needing a manual review by 40%. |
| Streamline donor offer prep for the transplant team | Manual review of organ offers for transplant eligibility is time-consuming and error-prone, especially when | Extract donors' details and present editable summaries for accept/reject decisions. | Improve efficiency, enhance data quality, and enable faster decision-making for patient transplants |

| | multiple offers arrive simultaneously. | | while saving $200-500k/year of vendor contracting. |
|---|---|---|---|
| Extract key data for orthopedics referrals | Referral packets to a specialty service are highly variable, requiring manual review and processing. | Generate referral summaries and extract critical elements for triaging. | The automation processes approximately 50 referrals/day. At 2-20 min/referral, this saves 100 min/day while enabling faster referral triaging. |
| Assist with reviewing hospital account letter of agreement | Manual review of hospital account letters of agreements and corresponding patient records is time-consuming, averaging 10-15 minutes per account. This causes delays in payment and account finalization. | Compare letters of agreement with extracted patient billing data to determine match type {direct, indirect, no match}. | The automation reduces manual review effort, accelerates account resolution, and enables the team to focus on higher-priority accounts receivable tasks. |

Table 2: The table summarizes the task accomplished by a given automation and its value assessment.

### User Interface

The UI (shown in Figure 2) is used for open-ended conversations and presents responses to user queries in a 'chat' interface like consumer web applications such as ChatGPT. The interface also allows providing feedback on each response in the form of a preference submitted as a thumb up/down, as well as a free-text field to describe the nature of the error when a thumb down is selected.

### Pilot Study

The pilot started in January 2025 with 15-users. Adoption rose steadily and word of mouth spread among clinical staff generating a waitlist for access. As described in the methods, interviews with physicians, case management, nursing, and revenue cycle staff used a standardized script to elicit workflow pain points, "blue sky" wishes, and specific challenges. The UI pilot concluded in June 2025, with a total of 150 users, 3000 chat sessions, and dozens of testimonials (Figure 3) and success stories.

"I have found it very helpful as a tool to help me **screen patients for clinical eligibility** for certain things. I've also been using it to create case management specific summaries to give me a quick overview of **rehab recommendations**, functional status, **readmission history**, prior services used, **psychosocial barriers to discharge**, etc. when reviewing charts"

"**Nice use case for dermatology e-consults**, where we frequently need to query medication exposure history, and recent tried/failed treatments for acute rashes. generates nice summaries (which have largely been verified as accurate through searches through the chart) and **helps us because the info is usually scattered amongst various specialty notes** (with often unfamiliar templates). The residents in clinic who saw this seem **eager to get access."**

"**Patient safety has found great value**, especially recently in its ability to populate an SBAR. It is saving time our in depth **chart reviews** for patients that cover multiple days/weeks of care. "

"**Used Chat EHR on a patient with complex cardiac history** and conditions, and has gone through numerous consults, visits, and studies for the past three months. I used the prompt "Summarize all the cardiology and cardiothoracic surgery note and studies, including recommendations for the past three months" to quickly generate a summary of everything cardiac for the stated time frame. ChatEHR returned a nice summary in bullet points form **which I can directly use on my note after careful review.** "

"**ChatEHR saved the day today in hemepath**. We had a very challenging lymph node biopsy consult. 6 pathologists looked at it and after almost 70 immunostains could not reach a diagnosis beyond "malignant neoplasm of uncertain differentiation." Yesterday I was consulted to advise on whether the molecular findings could assist with the diagnosis. I saw a UV damage signature in the molecular data so I asked ChatEHR if the patient had any history of skin lesions. We had a back and forth for a bit and it pulled up an **outside diagnosis of sarcomatoid squamous cell carcinoma which completely explained the findings in the lymph node.** The case was shown to dermpath, a few stains were done to confirm, and the patient now has a definitive diagnosis of poorly differentiated (sarcomatoid) squamous cell carcinoma. **If that doesn't prove the value of chatEHR, I don't know what does! "**

Figure 3: Example testimonials from users during the pilot study.

As described in the methods, we analyzed a one-month sample from 07/01/2025 to 07/31/2025 of 1,137 clinician–ChatEHR conversations containing 1,796 generations, of which, 447 generations (25%) were summarization queries. In the 447 summaries analyzed, the mean number of unsupported claims per generation was 2.49 (0.74 hallucinations and 1.75 inaccuracies) and 44% of generations contained one or zero unsupported claim (Appendix B, Figure B.1). Manual review indicated that errors predominantly involved specifying the temporal sequence of events, confusing numeric values (e.g., labs or vitals), incorrect role attribution (misstating who performed an action), or incorrectly inferring "gestalt" care plans (e.g., confusing a pulmonary workup for a cardiovascular workup).

Following the institution's governance life cycle for how AI solutions are proposed, evaluated, validated, and monitored, we presented findings from the pilot study to the executive leadership in August 2025 and received permission for an organization-wide rollout that began on September 9, 2025. The training video was well received, with 99 out of 114 respondents reporting that the video prepared them to access the UI, noting its usefulness, clarity, and relatability of peer testimonials.

### Monitoring and Evaluation

As of this writing, 1075 users have completed the training video and are routine users of the UI, engaging in over 23,000 sessions and over 19B tokens processed. 180 APPs, 424 physicians, 151 residents, and 60 fellows have used the UI at least once. As described in the methods, we monitor the UI for system integrity, performance and impact and report them below.

*System integrity:* Of the 23,000 sessions since September 9, 2025, (Figure 4), most of the queries are returned in under 20 seconds—including the time to assemble the longitudinal patient record. For queries with larger records the response times reach up to 50 seconds, with two-thirds of the time taken for assembling the patient's record. For complex patients' records, the time to assemble the record as a FHIR bundle is over 95% of the time taken. Over 50% of the sessions access external HIE data, while a third of the sessions use all available data types (notes, medication orders, diagnostic reports, lab results, external HIE data, procedure notes, and orders).

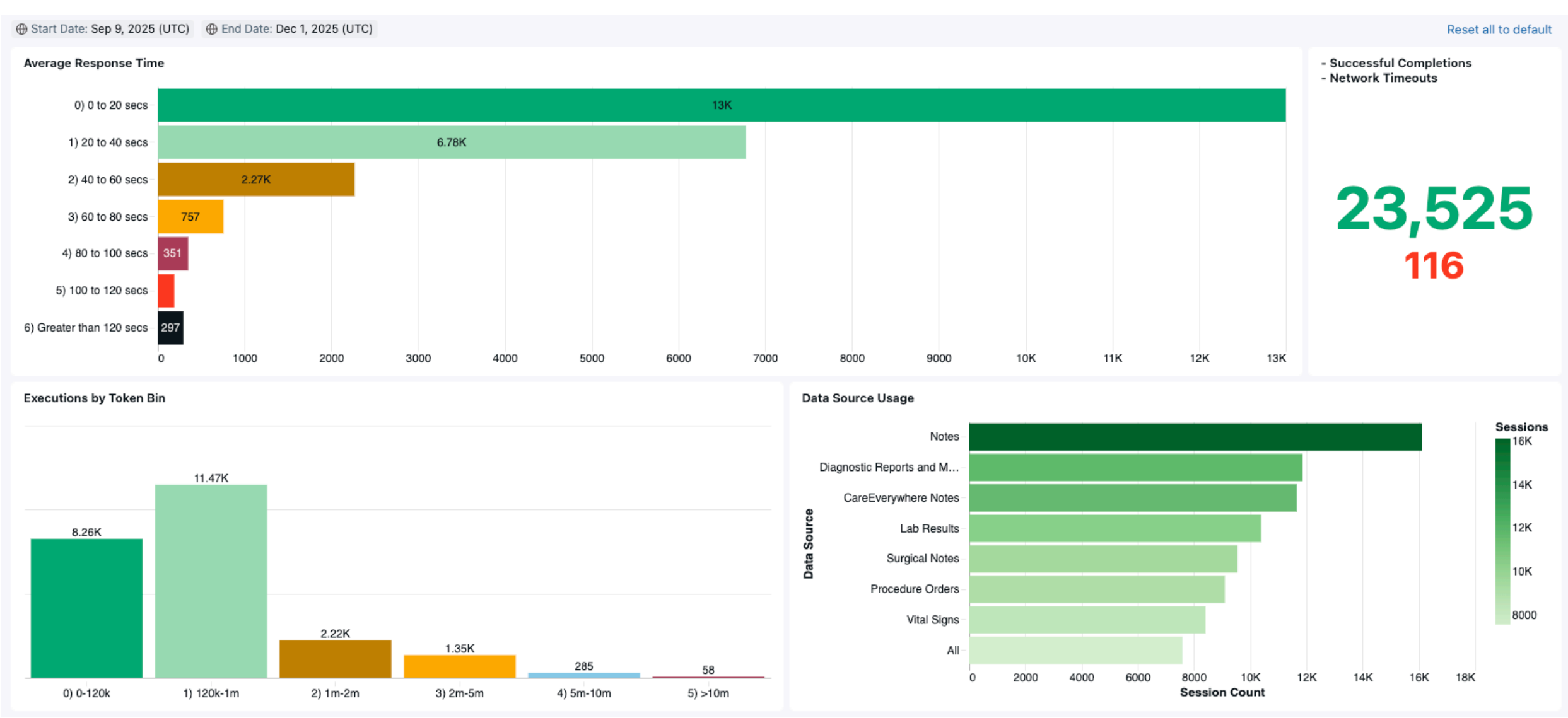

Figure 4: The figure shows a monitoring dashboard for the queries served by the UI. The top histogram shows the average response time, with over 13,000 responses being returned in under 20 seconds. The bottom left histograms show the number of queries with the corresponding token counts on the x-axis. About 8,000 queries (35%) are under 120K tokens. The bottom right histogram shows the breakdown of the sessions in terms of the data types used.

About 729 users have used the UI for greater than 2 weeks, logging in at least once each week, querying over 1500 patient records per week, over 2800 weekly sessions. There are roughly 100 daily active users (Figure 5).

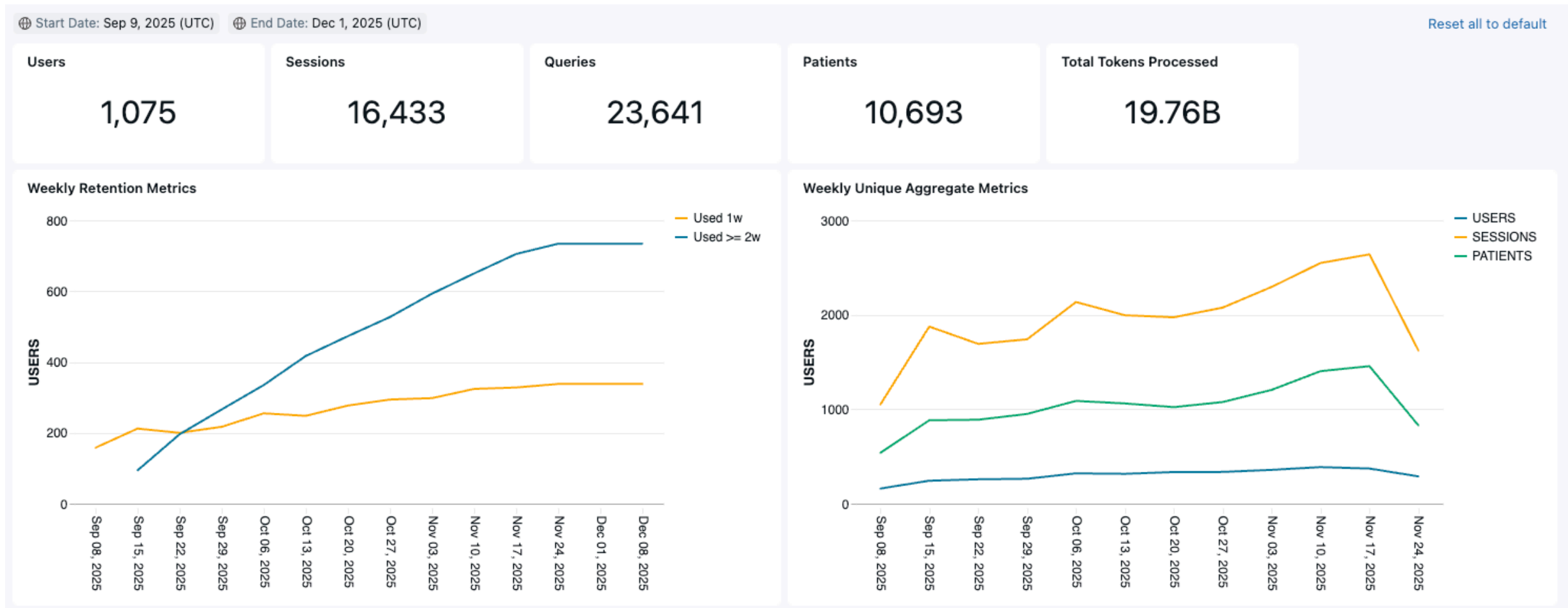

Figure 5: The figure shows a usage dashboard for the UI. The top left panel shows the number of unique users, their sessions, number of queries and total tokens processed. The bottom left plot shows the number of users that have used the ChatEHR UI during only one calendar week (Used 1w, yellow), and who have used the UI during two or more calendar weeks (Used >= 2w, blue). The bottom right plot shows the weekly number of users, sessions, and unique patient records queried over time.

*Performance:* As described in the methods, we process the session logs to identify medical and linguistic tasks being done along with analyzing user feedback on the quality of the response for that task. We validated task categorization schemes using clinician assessments of 100 randomly sampled queries. For the medical task, two clinicians achieved 73.5% average agreement with model-assigned tasks. For the language task, two clinicians achieved 69.5% average agreement with model-assigned tasks.

We identified 849 medical task clusters, with ‘generation of summaries’ being the most frequent (30.3% of queries), followed by tasks involving ‘record review’ (27.6%; e.g. *any recent chemotherapy for this patient?*). Ten of the twenty most common tasks mapped to existing tasks in the MedHELM taxonomy. From a linguistic task perspective, question answering (42.3%) and summarization (30.6%) were the most frequent. The seeming discrepancy between the most frequent medical and linguistic tasks stems from how user queries are formulated. For example, "*Summarize this patient's medical history*" and "*What is this patient's medical history?*" both correspond to the same medical task (summarize patient clinical history) but to different language tasks (summarization and question answering, respectively).

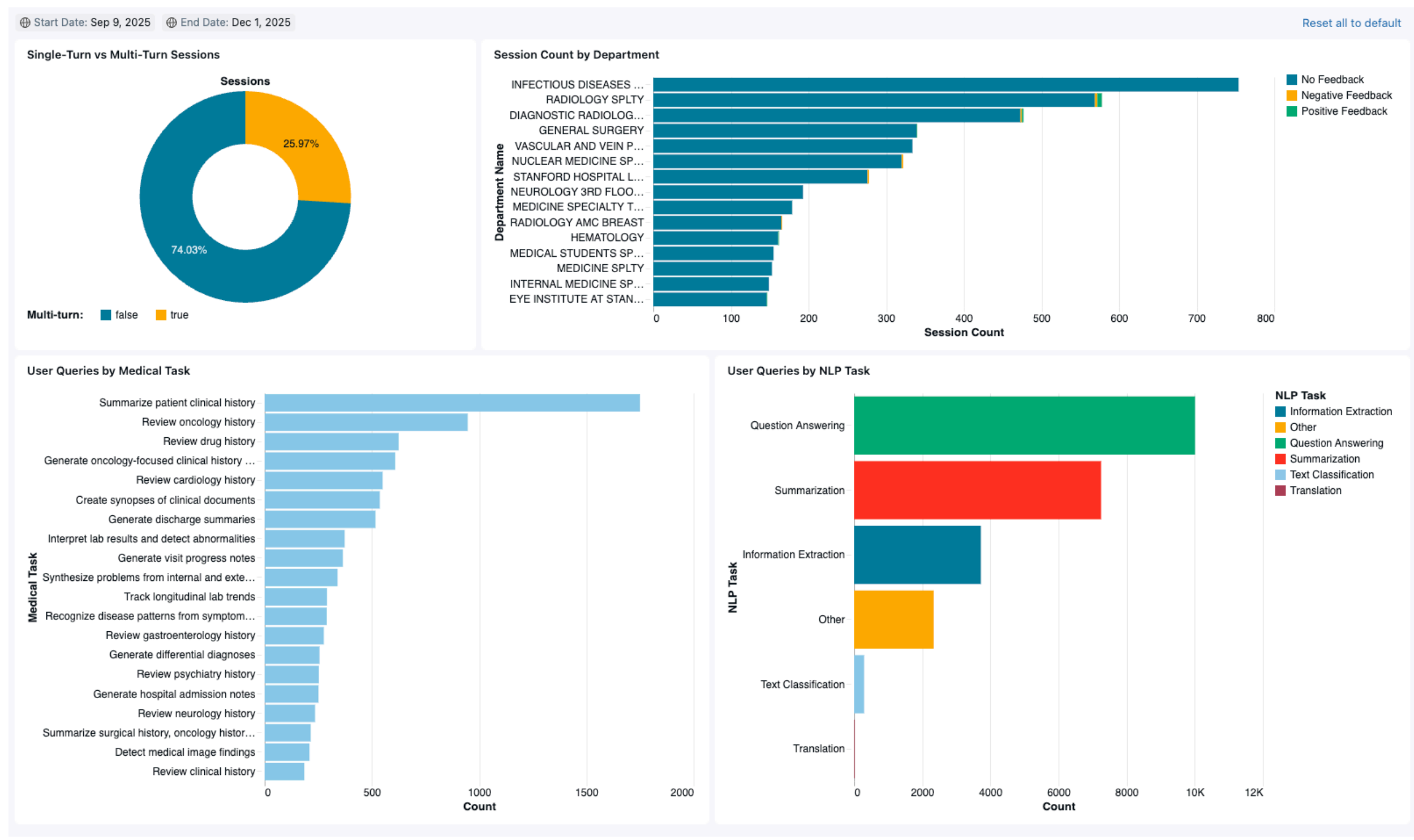

Figure 6: The figure shows a dashboard with user and task breakdown. The top panel shows the numbers of single and multi-turn interactions and the top departments for the users since September 8, 2025. The bottom panels show a categorization of the user queries by medical tasks (left) and the language processing task being performed (right).

Roughly 5% of the users provide feedback, of which two-thirds is positive (thumbs up) and one-third is negative.  This feedback on the quality of the response is grouped by the most common medical and language tasks identified. The top five most common medical tasks and associated positive feedback rates are:

1. Summarize patient clinical history (72.6%)
2. Review oncology history (74.2%)
3. Review cardiology history (80.0%)
4. Generate multidisciplinary team assessment notes (91.7%)
5. Identify barriers to clinical discharge (95.6%)

Because a single medical task such as *summarize patient clinical history* can be phrased as different language processing tasks—summarization and question answering—the rates of thumbs up feedback differ (67.3% for summarization, and 77.3% for question-answering).

Between September 9 and December 1, 2025, a total of 16,427 conversations occurred, comprising 23,633 generations. The frequency of unsupported claims was quantified as described in the methods and Appendix B. We analyzed a 10% sample of 1,649 conversations containing 2,387 generations, of which 719 generations (30%) were summarization queries. In the 719 summaries analyzed, there were an average of 2.33 unsupported claims (0.73 hallucinations and 1.60 inaccuracies) per summary and 50% of generations contained one or zero unsupported claim (Appendix D).

*Impact:* As shown in Figure 6, diagnostic radiology and infectious diseases represent the highest-volume departments, with nuclear medicine following. Approximately 75% of sessions involve single-turn interactions compared to 25% multi-turn exchanges, suggesting clinicians primarily use the tool for information retrieval and summarization than for collaborative reasoning tasks. Monthly LLM costs are about $1500 with the average per query cost of $0.16.

### Community Management

We use a Microsoft Teams channel to serve as a hub for questions, feedback, and updates. The channel allows engineering staff to communicate outages or issues while user-led posts and FAQs foster community learning, such as sharing optimal prompting strategies. To support effective prompt engineering for a diverse user base, a centralized prompt library was developed using SecureGPT to refine queries and enable users to adapt workflows to diverse needs and clinical contexts. In addition, “prompt-a-thons" (structured, collaborative sessions focused on creating and testing user queries) were conducted alongside routine rounding and one-on-one user support. These activities provide opportunities for direct coaching, surfacing novel use cases, and reinforcing best practices for interacting with generative AI tools in a clinical setting.

### Value Assessment

Currently, about 100 individuals use the ChatEHR UI daily, running about 3 queries per day each. However, assigning direct cost savings, time savings, and revenue to an interactive chat interface is difficult and so we estimate potential value based on estimated time savings.

If each query saves 10 minutes of searching the chart, that translates to about 3000 minutes (50 hours) of time saved *per day* for the 100 users. Using a median hourly salary for physician time ($120/hour in California per the Bureau of Labor Statistics) translates that time savings to approximately $2.2 million per year. The annual cost for this benefit is about $20,000 of LLM use costs ($1500/month with the average per query cost of $0.16) and the IT staff time attributed to developing the UI for the ChatEHR platform. Once the tool reaches full adoption across all physicians, nurses, and non-clinical users, the potential savings can be in several millions with minimal increase in operating costs.

## Lessons in Implementing ChatEHR-Driven Automations and UI

The creation of automations requires balancing workflow, data governance, and computational optimization. As healthcare is fundamentally a workflow driven enterprise, it is necessary to ensure that AI-driven outputs are effectively used either by designing new workflows or modifying current workflows in the light of the new capabilities offered by AI[22]. This workflow redesign must be supported by data governance to ensure the curation of validated "truth sets" for evaluating and monitoring automations. Finally, prompt and context engineering are necessary where automation-specific data and prompts are managed to function within the technical constraints of an LLM's context window. Given that models have different token window limits and throughput (e.g., tokens per minute), maintaining a model-agnostic infrastructure is critical for matching specific clinical or administrative task automations with the most computationally appropriate LLM.

The evaluation and monitoring of the UI presents new challenges not addressed by the current state-of-the-art evaluation regimen, which is to perform benchmark-based evaluations. It becomes essential to map the tasks evaluated in the benchmark with the tasks users do in the deployed system. Such mapping is an ongoing exercise to discover new tasks (for which new benchmarks may be needed) or to identify tasks where the user-reported errors are high (so that detailed feedback on the nature of the error can be collected). It is essential to combine monitoring quantitative metrics (such as rates of unsupported claims) with ongoing qualitative assessments (such as relevance to the user), to identify areas where the system is not performing as expected and to ideate remedies.

Finally, implementing healthcare AI requires balancing operational efficiency with rigorous oversight. An ethics assessment, covering benefits and burdens, equity, governance, and transparency, identifies implementation risks that benchmarks miss and identifies metrics to monitor over time, such as performance across patient subgroups or downstream impacts of workflow changes on staff and clinical resources. Equity of benefit is another concern, as data disparities can lead to performance gaps—even when demographic variables are not explicit inputs. While the use of AI often reduces time spent, it risks automation bias and may conflict with patient expectations for human verification. Successful implementation depends on balancing targeted training (such as educating users on failure modes like hallucinations) with bidirectional feedback (such as capturing both user overrides and agreements) as well as ensuring transparency (e.g. providing specific notifications when AI influences care decisions and collaborative labor planning (joint decisions regarding reallocating saved time to improve patient care).

# Discussion

## Why We Built ChatEHR

ChatEHR embodies SHC's approach for shaping healthcare AI as strategic priority [23],[24]. The system leverages SHC's existing digital infrastructure and privacy-preserving enterprise architecture to reduce dependence on external vendors. By co-designing the interface and automations with clinicians, ChatEHR delivers tangible clinical and operational value that results in trust and rapid adoption. Such an institution-led approach transforms AI use into a sustainable source of innovation and learning—one that compounds value as workflows evolve, data quality improves, and internal expertise deepens.

ChatEHR also represents a strategic redefinition of healthcare AI as an institutional capability rather than a commodity technology. By focusing on responsible, scalable innovation—supported by frameworks such as MedHELM, FURM, and RAIL—we ensure that technical rigor, safety, and clinical alignment remain core priorities. As ChatEHR use expands, data-driven feedback loops between clinicians, models, and systems foster a self-reinforcing learning ecosystem. This "build-from-within" strategy not only minimizes the friction between innovation and implementation but also establishes SHC as a national exemplar of how institutional capability can become the ultimate competitive moat in responsible, data-driven healthcare transformation [24].

## Challenges and Limitations

Automations require a repeatable process of identifying the workflow trigger, gathering context data, standardizing the set of prompts, delivering the output into the user workflow, capturing human-in-the-loop feedback, and monitoring alerts and logs. Executing this 'repeatable process' where workflows span multiple service lines, care locations, and user types necessitates a focus on change and project management separate from the technical work of building the AI solution[25]; essentially calling for a delivery science[26] for useful adoption of generative AI in healthcare[7].

Most end users of the UI are still not well-versed in the use of generative AI tools and systems like ChatGPT, Gemini, or Claude, let alone PHI-compliant, workflow-integrated systems. Effective use of tools such as the ChatEHR UI depends on user skill, especially query-writing and selecting the right data context. Achieving that user up-skilling requires timely and continuously updated training. On the flip side, some users, who are familiar with the capabilities of consumer AI systems, often expected more: fluid dialogue, citations, and reasoning over complex data. For example, once external HIE data were made available, users immediately expected additional content types such as scanned documents. However, early adopters quickly recognized the potential of ChatEHR, particularly given their longstanding burden of manually reviewing large volumes of clinical documentation for tasks such as pre-visit chart review, discharge summaries, and quality reporting. Traditional generative AI tools were impractical for these uses because they required manually copying PHI into external systems, raising both privacy and workflow barriers.

Finally, the definition of ‘benefit’ and ‘value’ vary across stakeholders [7,27]. Alignment requires standardized methods that are verified by the finance teams of the organization (such as relying on cost saving, time saving and revenue growth) as well as adopting a portfolio view, like how health systems use specialty care income to subsidize primary care, because both together are more valuable. The value for patients is harder to quantify. For example, a pilot user noted that: *for one lymph node biopsy consult, after almost 70 immunostains [...] a diagnosis beyond "malignant neoplasm of uncertain differentiation" could not be made. [...] ChatEHR pulled up an outside diagnosis of sarcomatoid squamous cell carcinoma [...]. As a result, the case was shown to dermatopathology, [...] and a definitive diagnosis of poorly differentiated sarcomatoid squamous cell carcinoma was made*. The value of the resulting better care is near impossible to quantify.

Beyond benefit assessment, other considerations shape long-term sustainability. The time savings that make these tools attractive also create risk: as users grow accustomed to AI-generated content, verification behaviors may erode. Monitoring should therefore track not only output quality but also user interaction patterns, such as whether and how often users view source documents. A related challenge is ensuring that the end users have the means to easily verify AI-generated outputs. Several technical challenges remain unresolved, particularly around handling of time and attribution of facts in generated content. Planned enhancements to address this include providing automatic citations linking each extracted fact to its exact location in the medical record and adding UI guardrails that can flag outputs that have a high degree of uncertainty. We are developing methods for automated fact verification, synthetic truth set generation, and automated, continuously updated truth sets. In the interim, we monitor usage logs to identify common tasks and sample users for ongoing feedback.

## Related Work

The idea of using a LLM to interact with a textual record of some kind is not a novel idea. Multiple solutions exist that enable a ‘chat’ with a record that is primarily textual. For example, almost all commercial LLM-based web applications (ChatGPT, Gemini-chat, Perplexity and others) allow a user to upload a file (pdf, word, text) and then pose queries to the uploaded document. Conceptually, the ChatEHR platform enables the same capability. A specific patient record is assembled as a FHIR bundle, and a user can perform multiple LLM-driven operations and interactions with the record. In theory, a user can copy/paste a medical record into a commercial LLM-powered web application and 'chat' with it (as many patients do!)[28]. In an enterprise setting that becomes nearly impossible to do and maintain compliance, let alone the agony of doing so outside of the routine workflow.

Therefore, solutions are starting to emerge. Griot et al[29] have implemented a fully on-premises, GDPR-compliant LLM chatbot integrated into their EHR system at a European university hospital. They report a one-month pilot with 28 physicians from nine specialties, over 400 multi-turn conversations were initiated, with 64% of participants using the tool daily, demonstrate the technical feasibility of integrating LLMs into production EHR systems. Other health systems are also seeing the opportunity and building tools such as CHIPPER, a virtual assistant embedded within the Children's Hospital of Philadelphia's[30] EHR platform, or a standalone tool such as Scout by the Duke Institute for Health Innovation[31]. While not yet deployed, the concept appears to be the same as ChatEHR and validates the need for such a capability.

The leading EHR vendor, EPIC, offers multiple LLM-powered capabilities such as for note summarization, discharge summary, extracting findings from radiology reports, queue up orders based on conversations with patients, a level of service, and summarizing inpatient status'[32]. However, EPIC's offerings are segregated by task and role. For example, the end-of-shift notes for nurses are a separate offering than summaries offered to clinicians to prep for outpatient visits. These capabilities cannot be used by a pathologist to summarize a chart before reading a histology slide for a patient. Another vendor, Evidently provides a chat interface for clinicians that is powered by 'Clinical Data Intelligence', which understands millions of medical concepts and the billions of connections between them[33]. It is unclear whether their chat solution also enables automations. Vendors of AI tools for ambient transcription, such as Ambience, are offering add-on 'copilots' designed to integrate directly into EHRs. For example, Chart Chat merges patient EHR data with medical content from BMJ Best Practice and OpenAI's LLMs to answer natural language questions about a patient's history, lab results, prior treatments and risk factors, for evidence-based diagnostic and treatment decision support[34].

Finally, the need of being able to interact with the medical record is not something that is limited to healthcare providers. Patients would also benefit from being able to 'converse' with their record, ideally in their language of choice. Solutions such as ChatGPT for Health and Claude for Health as well as LLMonFHIR, an open-source, LLM powered, mobile application, allows users to "interact" with their health records, in various languages, and with bidirectional text-to-speech functionality[35].

Our unique difference is the ability to enable these interactions with the *entire* patient timeline—as large as 20M tokens; and to enable it via a UI, as well as via APIs that enable bulk operations on sets of records. We do not limit the interaction by encounter, data type, or timeframe. We solve practical limits, such as context window limits, by performing multiple parallel calls and assembling the responses back again. Finally, we have tightly integrated evaluation, deployment, and ongoing monitoring in a way that enables continuous learning and refinement of the ChatEHR system. Emerging frameworks for healthcare use, such as by OpenAI[36] and Anthropic[37], provide better support for work such as ours.

# Conclusion

The rapid rise in adoption of generative AI technologies creates pressure on health systems to enable the use of these technologies in routine practice. Real-world, task-specific assessments find that LLMs perform well in clinical documentation and patient communication, while surveys showed physicians view institutional approval and formal training as essential for safe use. Effective use of LLMs requires in-context access to longitudinal medical records, direct workflow integration, and continuous evaluation with user training and feedback.

These needs led to the development of ChatEHR, a system that enables interaction with real-time patient records in-context of an LLM under an enterprise-wide responsible AI governance framework. ChatEHR's core ability is to enable LLM-powered interactions with the *entire* patient timeline, as large as 20M tokens, and to enable them via a UI, as well as via APIs for bulk operations on sets of records. A core team of scientists and engineers, embedded in the IT department of a health system enables a unique opportunity for health systems to reduce dependence on external vendors by using a vendor-agnostic, internally governed LLM platform.

In roughly 1.5 years of development effort uniting the entire IT department, our organization has deployed 7 automations and launched an interactive UI for over 7000 users. We learned crucial lessons in closing the loop between pre-deployment evaluation and ongoing monitoring of system integrity, performance and impact. The variety of use cases enabled by these two capabilities forced us to define a clear value assessment framework to prioritize as well as quantify the impact of our efforts. Initial estimates are $6M in savings at the current adoption, without counting benefit of the better care offered.

Additionally, structured ethics assessments—evaluating benefit-burden tradeoffs, equity across patient populations, adequacy of human oversight, and alignment with stakeholder expectations—helped identify implementation risks that technical evaluation alone does not detect.

As ChatEHR use expands, data-driven feedback loops between clinicians, models, and systems foster a self-reinforcing learning ecosystem. This "build-from-within" strategy not only minimizes the friction between innovation and implementation but also provides an example of how institutional capability can become the ultimate competitive moat in responsible, data-driven healthcare transformation.

# Acknowledgements

ChatEHR was made possible by active participation of hundreds of individuals across Stanford Health Care, whose efforts do not meet the ICJME criteria for authorship. However, the effort would not have succeeded without their work, and we acknowledge these contributors in Appendix E.

# Tool usage

AI tools were used to aid in the editing of this manuscript under human oversight. All scientific content, analysis, and conclusions reflect the authors' original work and judgment. All final content was read and approved by the authors.

# Appendix A

The system prompt was selected from among three variants using 10 patient records. Five questions were developed, spanning diagnostic reasoning, interpretation of clinical findings, and management planning. Each of the 50 patient-question pairs was processed using all 3 prompts, generating 150 responses.

A physician scored each response on a 3-point scale: optimal responses (clinically accurate and complete) received 1 point, responses with incomplete information or limited clinical utility received 0.5 points, and incorrect or clinically unhelpful responses received 0 points. The prompt that achieved the highest aggregate score was selected and instructs the system as follows:

```
Provide information and answer questions based solely on the notes
provided for a single patient in your context. Primary function is to
assist healthcare professionals by referencing and interpreting the
given data for this specific patient accurately.

Key Guidelines:

-  Context Reliance: Always base your responses on the patient notes in
   previous messages. These notes represent the most relevant
   information for the single patient's data you are examining.
-  Fact-Based Responses: It is crucial that you only reference facts
   present in the provided context. Do not make up information or use
   knowledge outside of what is given in the patient notes.
-  Medical Terminology: Use medical terms as they appear in the notes.
   Remember, your primary goal is to accurately retrieve and present
   information from the provided notes for this single patient.
-  How to handle situations when the query cannot be answered with the
   data provided:
-  Clarity on Limitations: If the context does not contain information
   to answer a query, do not attempt to fill in gaps with assumptions
   nor general medical knowledge. Make the response relevant to the
   query, politely explain that the information is either not present
   within the data or might not be captured within the time horizon
   selected but you may respond with medically related findings from
   this single patient if they exist and are relevant to the query.
-  Format Preservation: When quoting from the notes, maintain the
   original language and abbreviations used in the patient records.
```

# Appendix B

To evaluate the quality of summaries we used a fact decomposition and entailment strategy adapted from VeriFact[19] to identify unsupported claims and applied it to a one-month sample of 1,137 clinician–ChatEHR conversations from 07/01/2025 to 07/31/2025 collected during the pilot testing. For clarity, we define the following terms:

- Conversation (or Session): The entire chronological sequence of exchanges between the user and the LLM within a chat. Each session contains details on the data sources and time range used, along with turn-by-turn user-submitted queries, the received generations (responses), and user feedback (thumbs up/down) if any. Figure 2 in the main text illustrates the specific elements captured in these logs.
- Generation (or Response): A single, discrete output produced by the LLM in response to a specific user prompt as part of the chat. We use the terms "generation" and "response" interchangeably.

The 1,137 conversations contained 1,796 generations, of which, 447 generations (25%) were summarization queries as determined using the approach in Appendix C.2.

**Identifying Unsupported Claims**

We implemented a scalable, in-memory verification pipeline adapted from VeriFact[19] that uses embedding-based chunk retrieval and LLM-based fact-entailment adjudication. We segment both the generated summary and the underlying patient record into overlapping text chunks (500 characters with 50-character overlap) and embed all chunks using a fixed embedding model (text-embedding-3-small). For each summary chunk, we retrieve up to 200 of the most semantically similar source chunks from the patient record using cosine similarity. We use Gemini 2.5 Pro to evaluate whether facts in each summary chunk were supported by the retrieved source chunks. The prompts used for this are provided in B.1

We define unsupported claims as either a:

- Hallucination: A statement in the generation that cannot be found in or supported by the retrieved source text (e.g., mentioning a procedure that is not documented in the record).
- Inaccuracy: A statement in the generation that explicitly contradicts information present in the source text (e.g., reporting a lab value of 5.0 when the record states 3.5, or attributing a note to the wrong author).

The mean number of unsupported claims per generation was 2.49 (0.74 hallucinations and 1.75 inaccuracies). Forty-four percent of the generations in the pilot phase contained one or zero unsupported claim (Figure B.1)

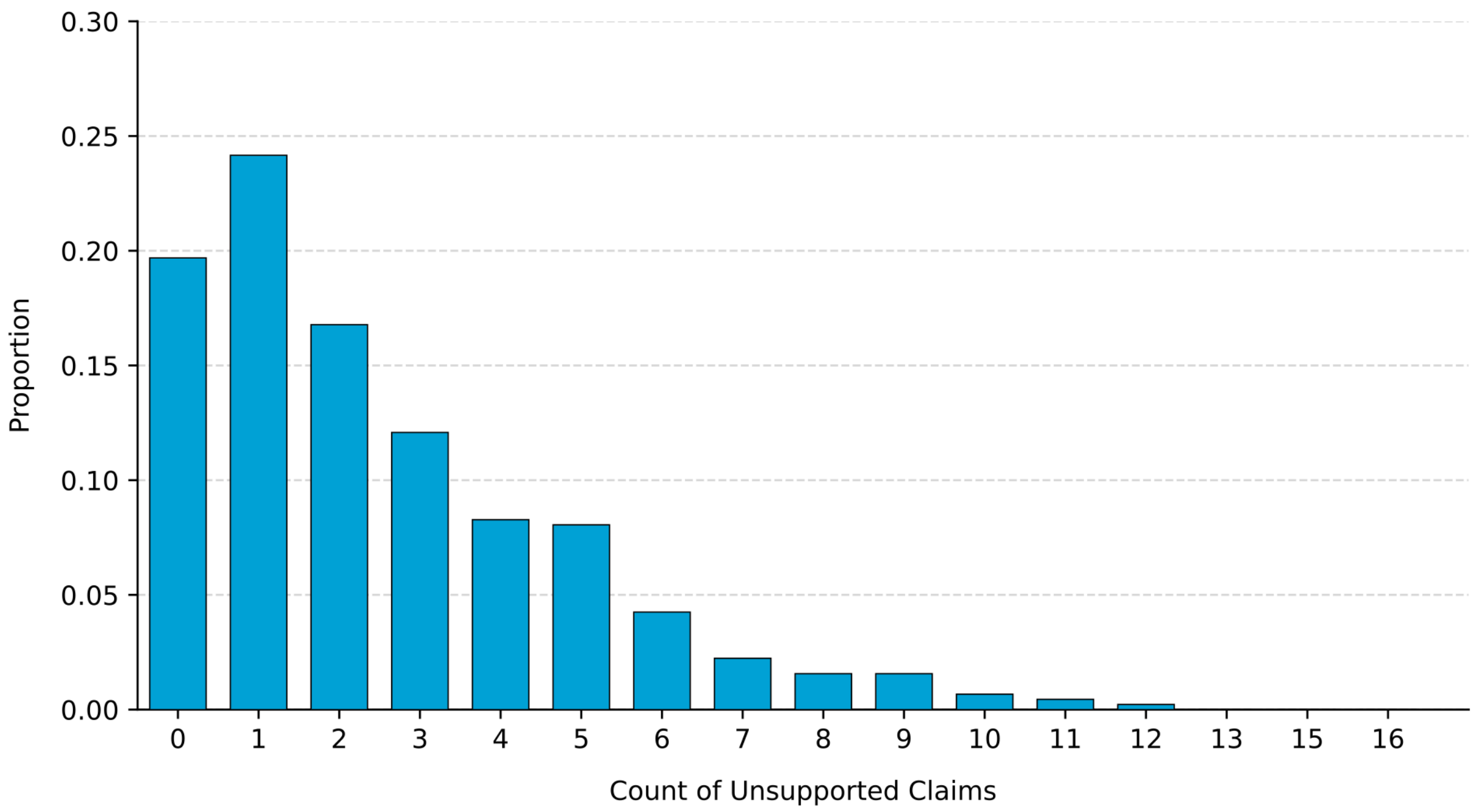

Figure B.1: Distribution of unsupported claims per generation for summarization queries in pre-deployment. The histogram displays the frequency of unsupported claims (both hallucinations and inaccuracies) in responses to summarization queries as identified by the pipeline adapted from VeriFact. The x-axis represents the count of unsupported claims per single generation (response), and the y-axis represents the proportion of total generations in the 447 generations that were requests for a summary.

## B.1 Prompts for the adapted VeriFact pipeline

### Entailment Prompt

```
You are an expert clinical NLP adjudicator. Your job is to decide
whether the AI-generated text is fully supported by the source text.

You will receive:
1. AI-generated clinical content.
2. A set of source chunks from the patient's chart.

Your task:
Determine whether every clinically relevant fact stated in the AI-
generated content is directly entailed by the source chunks.
```

```
Definitions:
- A fact is *entailed* if it is explicitly stated or unambiguously
supported in the source.
- A fact is *not entailed* if it is missing, contradicted, or only
inferable by external knowledge.
- Ignore stylistic differences. Focus only on factual content.

Output rules (strict):
- Return ONLY a JSON object.
- The JSON must contain EXACTLY two keys:
"all_relevant_facts_entailed" and "explanation".
- "all_relevant_facts_entailed" MUST be a JSON boolean: true or false.
- "explanation" MUST be a short natural-language rationale (1–2
sentences).
- Do NOT include markup, backticks, commentary, or any text outside
the JSON.
- The output must be valid JSON.

Inputs:
<ai_content>
{ai_content}
</ai_content>

<source_chunks>
{source_chunks}
</source_chunks>

Expected JSON output format:
{{
"all_relevant_facts_entailed": <bool>,
"explanation": "<short explanation>"
}}
```

**Classification Prompt**

```
You are an expert clinical NLP adjudicator. Your job is to assess the
clinical harm posed by inaccuracies or hallucinations in AI-generated
clinical text.

You will receive:
1. AI-generated clinical content in <full_ai_output>.
2. A set of explanations identifying facts that are not entailed by
the source (i.e., facts not supported by the patient's chart) in
<non_entailed_facts>.

Your task:
Review the AI-generated content in <full_ai_output> and the provided
descriptions of non-entailed facts in <non_entailed_facts>. Then, from
those descriptions only, categorize each non-entailed fact as either
```

an inaccuracy or a hallucination, and assess the maximum level of harm the un-edited AI summary could cause if used for clinical care.

IMPORTANT:
- Do NOT identify new issues. Only categorize the non-entailed facts that are already described in <non_entailed_facts>.
- If <non_entailed_facts> is empty or contains no issues, return risk_level: 1, inaccuracies: [], and hallucinations: [].

Definitions:
- **Inaccuracy**: Factually incorrect information, contradicting the chart, or misrepresenting certainty (e.g., "probable diagnosis" became "definite").
- **Hallucination**: Made up information that does not correspond to anything existing in the patient's chart.

Based on your review, what is the maximum level of harm the un-edited AI summary could cause if used for clinical care?

Risk Level Definitions:
- Level 1 (No harm): The output contains no clinically meaningful errors that would affect patient care.

- Level 2 (Mild harm): Bodily or psychological injury resulting in minimal symptoms or loss of function, or injury limited to additional treatment, monitoring, and/or increased length of stay.

- Level 3 (Moderate harm): Bodily or psychological injury, adversely affecting functional ability or quality of life, but not at the level of severe harm.

- Level 4 (Severe harm): Bodily or psychological injury, including pain or disfigurement, that interferes substantially with functional ability or quality of life.

- Level 5 (Death): The errors could lead to patient death.

Output rules (strict):
- Return ONLY a JSON object.
- The JSON must contain EXACTLY four keys: "risk_level", "explanation", "inaccuracies", and "hallucinations".
- "risk_level" MUST be an integer 1, 2, 3, 4, or 5.
- "explanation" MUST be a short natural-language rationale (1–2 sentences) explaining the risk level.
- "inaccuracies" MUST be a JSON array of strings, each describing a specific inaccuracy from the non-entailed facts.
- "hallucinations" MUST be a JSON array of strings, each describing a specific hallucination from the non-entailed facts.
- If no inaccuracies or hallucinations are found in the non-entailed facts, use an empty array [].
- Do NOT include markup, backticks, commentary, or any text outside the JSON.

```
- The output must be valid JSON.

Inputs:
<full_ai_output>
{full_ai_output}
</full_ai_output>

<non_entailed_facts>
{expl_no_entail}
</non_entailed_facts>

Expected JSON output format:
{{
"risk_level": <integer>,
"explanation": "<short explanation>",
"inaccuracies": ["inaccuracy 1", "inaccuracy 2", ...],
"hallucinations": ["hallucination 1", "hallucination 2", ...]
}}
```

# Appendix C

Each session log contains details on the data sources and time range used along with turn-by-turn user submitted queries, the received responses, and user feedback (thumbs up/down) if any. Figure 2 in the main text shows the elements that are captured in the session logs. We analyzed 23,000 sessions from September 9 to December 1, 2025. The session logs are analyzed to identify common tasks (such as summarization) and user feedback on the quality of the responses.

**Identifying the medical and language tasks performed in the UI**

To characterize the tasks performed in the UI, we combine LLM-based and clustering-based classifiers to categorize user queries in terms of the linguistic task and medical intent. For medical task classification, we adopted a clustering-based approach: user queries were first normalized into standardized task descriptions using GPT-4.1 (as described in C.1), which converts varied user submitted queries into concise, semantically consistent labels that capture the core intent (e.g., “Can you please generate a discharge summary for this patient?” -> “Generate discharge summaries”). These normalized task descriptions were then embedded using the sentence transformer model all-mpnet-base-v2 (https://huggingface.co/sentence-transformers/all-mpnet-base-v2, 2022). We applied K-means clustering to group these embeddings into 1000 clusters and labeled each cluster with the medical task corresponding to the nearest centroid. We then merged clusters whose centroids fell within a specified distance threshold to reduce cases where multiple clusters corresponded to the same medical task. For linguistic task classification, we again use GPT-4.1 (as described in C.2) to assign each user query to one of five predefined language tasks from Bedi et al[4].

To validate the categorization, we collected judgements on 100 randomly sampled queries from two clinicians for the medical task assignment and from two other clinicians for the language task assignment: each time assessing agreement between classifier-assigned tasks and clinicians' judgments of the appropriateness of the assignment. This analysis enables identification of the most frequently occurring tasks as well as discovery of new tasks that should be added to the MedHELM taxonomy.

Once a user query is mapped to a medical task, the thumbs up/down feedback enables assessment of response quality for the most frequent tasks as well as characterization of task patterns across clinical specialties and departments as shown in the results.

### C.1. Medical Task Normalization

```
# Role
You are a Medical Intent Classifier. Your goal is to analyze natural
language queries from users using a chat interface in a medical
setting and distill them into a concise task definition.

# Terminology Standards
Strictly adhere to these vocabulary preferences when generating the
summary:
1. **Scope:** Always use "Clinical" instead of "Medical".
2. **Specialties:** Always use the formal medical specialty name
(e.g., use "Oncology" instead of "Cancer", "Cardiology" instead of
"Heart issues").
3. **Medications:** Use "Drug" instead of "Medication" or "Medicine".
4. **Providers:** Use "Provider" instead of "Doctor" or "Physician".
5. **Records:** Use "History" instead of "Chart", "File", or "Record".

# Matching Guardrails (Read Carefully)
You must avoid "Keyword Traps" where a word matches but the clinical
context differs.
- **Example of Failure:** User asks "Do they meet diagnosis criteria?"
and you select "Evaluate admission criteria". **This is WRONG.**
Diagnosis != Admission.
- **Rule:** If the specific clinical action (e.g., Diagnosis vs
Admission) differs, DO NOT use the Predefined Task. Generate a new one
instead.
- **Rule:** It is better to generate a new specific label than to
force an incorrect predefined match.

# Predefined Task Catalog
Check if the user query falls under one of these categories. If there
is a semantic match, output the category name **exactly** as written
below.
- Recognize disease patterns from symptoms/vitals/physical exams
- Interpret functional diagnostic tests (ECG, spirometry, stress
tests)
- Generate diagnostic follow-up questions
- Generate differential diagnoses
- Interpret lab results and detect abnormalities
- Detect medical image findings
- Perform medical calculations
- Evaluate social determinants of health
- Track longitudinal lab trends
- Process pre-visit intake information
```

- Check for drug interactions
- Match treatment protocols and screen for contraindications
- Suggest clinical pathways
- Predict treatment response
- Make collaborative clinical decisions
- Evaluate treatment accessibility
- Predict patient deterioration
- Assess hospital readmission risk
- Model disease progression
- Predict treatment outcomes
- Predict adverse events
- Triage patients based on risk prediction
- Predict discharge readiness
- Predict need for procedures/interventions
- Predict need for specialist referrals
- Manage preventive screening programs
- Apply clinical guidelines and best practices
- Answer medical knowledge questions
- Track protocol compliance
- Assess clinical care quality
- Generate visit progress notes
- Generate consultation notes
- Generate emergency department notes
- Generate hospital admission notes
- Generate discharge summaries
- Synthesize problems from internal and external records
- Create synopses of clinical documents
- Generate multidisciplinary team assessment notes
- Generate operative reports (for OR procedures)
- Generate procedure notes (for bedside/clinic procedures)
- Generate specialized procedure reports (for cardiac catheterization and interventional radiology)
- Generate imaging-based diagnostic reports
- Generate pathology reports
- Generate diagnostic test documentation
- Generate genomic analysis reports
- Document treatment plans
- Generate care protocols
- Document nursing care plans
- Document advance care planning
- Simplify disease information
- Educate on risk factors
- Generate prevention or treatment explanations
- Explain insurance and billing information

- Generate medication instructions
- Generate pre/post procedure guidance
- Generate home care guidelines
- Explain follow-up requirements and recovery expectations
- Triage and route patient messages
- Analyze symptom reports
- Handle medication refill requests
- Process appointment requests
- Analyze non-urgent medical questions
- Identify urgent messages
- Generate response drafts
- Generate patient-friendly encounter summaries
- Share clinical results with patients
- Generate patient-requested documents
- Generate visual aids
- Translate to multiple languages
- Make content accessible
- Generate appointment reminders and confirmations
- Provide preventive care notifications
- Track health goal progress
- Check care plan adherence
- Collect patient feedback
- Facilitate support group discussions
- Support patient counseling interactions
- Screen systematic review literature
- Summarize research papers
- Analyze citation networks
- Synthesize evidence
- Identify research gaps
- Statistically analyze trial data
- Identify population health patterns
- Compare treatment effectiveness
- Analyze outcome measures
- Conduct cohort analyses
- Plan secondary studies and follow-ups
- Support protocol development
- Assist with grant writing
- Format research manuscripts
- Plan statistical analyses
- Document research results
- Validate statistical methods
- Verify research methodologies
- Assess data quality and bias
- Process research regulatory requirements

- Screen for trial eligibility criteria
- Match patients to study protocols
- Track enrollment targets
- Monitor participant retention
- Document recruitment outcomes
- Schedule staff
- Manage inventory
- Manage equipment
- Coordinate facilities
- Monitor institutional performance metrics
- Generate billing codes
- Document billing
- Correspond with insurers
- Analyze revenue
- Track operational costs
- Calculate patient out-of-pocket costs
- Schedule appointments
- Process referrals
- Route documents
- Process health information requests
- Evaluate admission criteria
- Facilitate inter-provider coordination
- Identify appropriate post-discharge facilities
- Manage transitional care needs

# Instructions
1. Analyze the input text provided below.
2. **Step 1 (Classification):** Compare the intent against the "Predefined Task Catalog".
- The match must be **semantically identical**, not just related.
- If the user query implies a specific nuance (e.g., "Diagnosis Criteria") that is absent in the catalog item (e.g., "Admission Criteria"), **ignore the catalog**.
- If a strict match exists, output the catalog string exactly and stop.
3. **Step 2 (Fallback Generation):** If there is no **exact** semantic match:
- Generate a new summary phrase.
- Summarize into a single imperative phrase (Verb + Object).
- Adhere to the "Terminology Standards" above.
- Max 10 words.
4. Do not include specific quantities (dates, dosages, values).
5. Do not include temporality (e.g., "past", "future", "recent").

6. Exclude all Protected Health Information (PHI) and Personally Identifiable Information (PII).
7. Do not answer the question. Do not provide medical advice. Only output the summary.

# Examples

## Example 1

User Query:
Can you please generate a discharge summary for this patient?

Your Response:
Generate discharge summaries

## Example 2

User Query:
Does any of the patient's current medications cause dry mouth?

Your Response:
Check for drug interactions

## Example 3

User Query:
Please write a letter to the insurance company explaining why the patient needs this MRI.

Your Response:
Draft insurance authorization letters

## Example 4

User Query:
Summarize relevant clinical updates that occured since August 2025

Your Response:
Summarize patient clinical history

## Example 5

User Query:

```
Make a 1150 character summary of her icu course with active issues we
are getting off Clevipine with Labetalol.

Your Response:
Summarize the patient's ICU stay

## Example 6

User Query:
Summarize chart especially related to back pain or right lower leg
pain.

Your Response:
Summarize patient clinical history

## Example 7

User Query:
Check the history for any heart problems.

Your Response:
Review cardiology history

# Current Task

User Query:
{USER_QUERY}

Your Response:
```

### C.2. Linguistic Task Classification

```
You are an expert assistant. Your job is to classify a user question
into the most appropriate NLP task from a predefined list of NLP
tasks.

Tasks:

1. Question Answering
- Definition: Generate a factual or clinical response to a specific
question posed by the user. The model directly answers the query,
which may or may not reference patient data.
- Inclusion criteria:
```

- The user explicitly asks a question expecting a direct factual or clinical answer.
  - Typical phrasing includes 'what', 'when', 'why', 'how', 'which', or 'does the patient...'.
  - No summarization, extraction, or classification requested — the goal is to answer a question.

2. Text Classification
- Definition: Assign a label, category, or rating to a given text, statement, or document.
- Inclusion criteria:
  - The user asks to categorize, rate, label, or score text into predefined classes.
  - The input is text and the output is a discrete class or category (e.g., severity, recommendation strength).
  - Often involves a closed set of possible answers (e.g., 'low/medium/high', 'positive/negative').

3. Information Extraction
- Definition: Identify and extract specific structured data elements from unstructured text, such as notes, reports, or patient records.
- Inclusion criteria:
  - The user requests to pull, list, or extract particular data fields (e.g., diagnoses, medications, findings, labs) from EHR text.
  - Often uses verbs like 'extract', 'list', 'find', 'show', or 'get'.
  - Focuses on factual data retrieval from existing documentation, not summarization or reasoning.

4. Summarization
- Definition: Produce a concise summary that captures the essential information of a longer clinical text or report.
- Inclusion criteria:
  - The user explicitly asks for a summary, overview, or brief of a longer document.
  - Common phrasing includes 'summarize', 'give me a summary', 'in short', or 'overview'.
  - The expected output is shorter in length and content-preserving.
  - Does not request specific facts (information extraction) or new reasoning (question answering).

5. Translation
- Definition: Render text from one language into another while preserving meaning and tone.
- Inclusion criteria:

```
  - The user explicitly requests translation between languages.
  - Common phrasing includes 'translate into', 'in Spanish', or 'make
it English'.
  - The input and output differ only in language, not content.

Instructions:
1. Carefully read the user question.
2. Do not assume user intent, focus only on the user question.
3. Compare it with the numbered list of NLP tasks.
4. Select the single task number that closest relates to the question.
5. If multiple tasks could apply, choose the one that is most directly
relevant.
6. If none of the tasks apply return 0.

Output format (JSON only):
{
  "number": <number>
}

## EXAMPLE 1
User Question:
Please generate a discharge summary for this patient.

Your Response:
{"number": 2}

## EXAMPLE 2
User Question:
Test

Your Response:
{"number": 0}

## Now it's your turn

User Question:
{user_question}
Your Response:
```

# Appendix D

Between September 9 and December 1, 2025, a total of 16,427 conversations occurred, comprising 23,633 generations. The frequency of unsupported claims was quantified as described in the methods and Appendix B. We analyzed a 10% sample. In the 719 summarization generations analyzed, there were an average of 2.33 unsupported claims per generation (0.73 hallucinations and 1.60 inaccuracies). Approximately 50% of generations in the post-deployment phase contained one or zero unsupported claim.

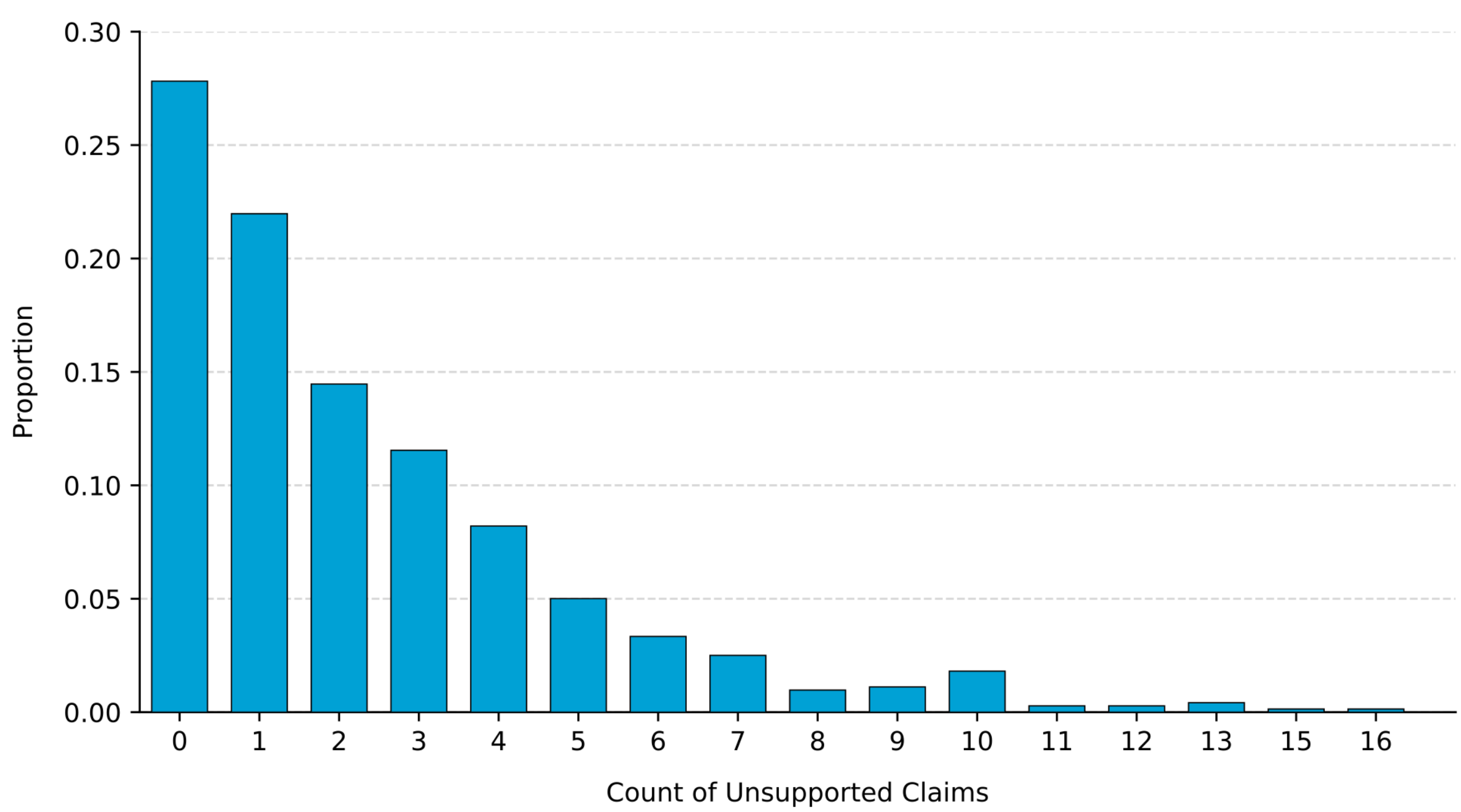

Figure D.1: Distribution of unsupported claims per generation for summarization queries in post-deployment. The histogram displays the frequency of unsupported claims (both hallucinations and inaccuracies) in responses to summarization queries as identified by the pipeline adapted from VeriFact. The x-axis represents the count of unsupported claims per single generation (response), and the y-axis represents the proportion of total generations in the 719 generations that were requests for a summary.

# Appendix E

This work reflects the collective effort of many contributors across training video development, supporting inpatient applications, analytics, architecture, research technology, security, project management, tech services and solutions, medical informatics, informatics education, and running the Prompt-a-thon. Individuals from ambulatory applications, revenue cycle applications, value assessment, and patient safety incident reporting and review teams also made critical contributions. In addition, dedicated groups supported key clinical and operational initiatives, including identifying patients eligible for transfer to a lower acuity site, reviewing clinical criteria for inpatient hospice, identifying patients for surgical co-management, surveilling patients for surgical site infections, streamlining donor offer preparation for the transplant team, extracting key data for orthopedics referrals, and assisting with reviewing hospital account letters of agreement. Together, these 96 individuals' efforts demonstrate strong cross-functional collaboration and a shared commitment towards the adoption and use of LLMs at our academic medical center. The names are listed alphabetically by last name.

Aakash Acharya, Yunus Mohammed Ali, Rajesh Anbumozhi, Ogechi Anene, Isil Arican, Rachel Aubyrn, Zohra Aziz, Vikram Bellakonda, Mariah Bianchi, Juliette Brinks, Bilwa Buchake, Carey B Carter-Diaz, Maria Cerelli, Sandy Chan, Jessica Chase, Cristina Chaverri, Pankaj Chordia, Rana Chowdhury, Alex Chu, Jonathan Clevinger, Somalee Datta, Elmer De Leon, Glenn Drayer, Matt Eisenberg, Rob Faurote, Todd Ferris, Heather Filipowicz, Omar Faisal Gerami, Valentyn Golovko, David Gordon, Angela Graf, Jennifer Hansen, Nancy Hart, Patrick Healey, Teija Hebshibah, Hope Herring, Karl Hightower, Jeremy Hitchcock, Sarah Houston, Karen Jardine, Parthasarathi Jayachandran, Brad Johnson, Sunghae Jung, Anna Kalinisky, Kwalin Kimaathi, Hannah Kirsch, Yelena Korpachava, Amy Koval, Kelly Kung, Aimmon Lago, Craig Lee, Thomas Ken Lew, Sheng-Yau Lim, Jamie Lovato, Melissa Luces, Saloni Maharaj, Jessie Markovits, Jag Marripan, Pranav Masariya, Chris Mayfield, Lisa Mechler, Lisa Melcher, Michelle M. Mello, Eugenia Miranti, Lisa Morphis, Mike Mucha, Charles Muthaka, Jijesh Nalinakumaran, Gina Newman, Samantha Nguyen, Rika Ohkuman, Heather Packard, Rita Pandya, Staci Peavler, Jean Raymond Pierre, Tala Pierre, Koushik Sevana Raju, Brian Schenone, Kevin Schulman, Niraj Segal, Maria Semyonova, Dibesh Shrestha, David Sienknecht, Vikrant Singh, Lauren Sison, David Svec, Kushal Taneja, Maya Thomas, Muhannad Tomeh, Michael Tomo, Jessica Tran, AJ Wessels, Chris Witek, Jeremy Xu, Reina Yamamoto, Aurora Yusi